\documentclass[sigconf]{acmart} 
\AtBeginDocument{%
  \providecommand\BibTeX{{%
    \normalfont B\kern-0.5em{\scshape i\kern-0.25em b}\kern-0.8em\TeX}}}

\copyrightyear{2025}
\acmYear{2025}
\setcopyright{acmlicensed}
\acmConference[]{}
\acmBooktitle{}
\acmPrice{15.00}

\usepackage{amsfonts, amsthm} 
\usepackage{algorithm}
\usepackage{algorithmic}
\usepackage{multirow}
\usepackage{subcaption}
\usepackage{textcomp}
\usepackage{longtable}
\usepackage{xcolor}
\usepackage{siunitx}
\usepackage[inline, shortlabels]{enumitem}
\usepackage{xcolor}
\usepackage{listings}
\definecolor{mGreen}{rgb}{0,0.6,0}
\definecolor{mGray}{rgb}{0.5,0.5,0.5}
\definecolor{mPurple}{rgb}{0.58,0,0.82}
\definecolor{backgroundColour}{rgb}{0.95,0.95,0.92}
\def\BibTeX{{\rm B\kern-.05em{\sc i\kern-.025em b}\kern-.08em
    T\kern-.1667em\lower.7ex\hbox{E}\kern-.125emX}}

\lstdefinestyle{CStyle}{
    backgroundcolor=\color{backgroundColour},   
    commentstyle=\color{mGreen},
    keywordstyle=\color{magenta},
    numberstyle=\tiny\color{mGray},
    stringstyle=\color{mPurple},
    basicstyle=\footnotesize,
    breakatwhitespace=false,         
    breaklines=true,                 
    captionpos=b,                    
    keepspaces=true,                 
    numbers=left,                    
    numbersep=5pt,                  
    showspaces=false,                
    showstringspaces=false,
    showtabs=false,                  
    tabsize=2,
    language=C
}
\lstdefinestyle{highlight}{
    morekeywords={@highlight},
    keywordstyle=\color{red}\bfseries, 
}

\lstdefinestyle{LLVMStyle}{
  morekeywords={
    define, declare, global, constant, internal, external, align, section, 
    module, target, datalayout, triple, 
    i1, i8, i16, i32, i64, float, double, void, 
    ret, br, switch, indirectbr, invoke, call, alloca, load, store, fence,
    cmpxchg, atomicrmw, getelementptr, trunc, zext, sext, fptrunc, fpext,
    uitofp, sitofp, fptoui, fptosi, inttoptr, ptrtoint, bitcast, addrspacecast,
    icmp, fcmp, phi, select, freeze, va_arg, landingpad, catchpad, cleanupret,
    catchret, catchswitch, resume, unreachable, 
    add, fadd, sub, fsub, mul, fmul, udiv, sdiv, fdiv, urem, srem, frem,
    shl, lshr, ashr, and, or, xor, 
    extractelement, insertelement, shufflevector, 
    dbg, !dbg
  },
  sensitive=true,
  morecomment=[l]{;},
  morestring=[b]",
  morestring=[b]',
  keywordstyle=\color{blue}\bfseries,
  commentstyle=\color{gray}\itshape,
  stringstyle=\color{orange}\ttfamily,
  basicstyle=\ttfamily,
  breaklines=true,
  columns=fullflexible,
  literate={\%}{\%}{1}
}
\title{PrETi: Predicting Execution Time in Early Stage with LLVM and Machine Learning}

\begin{document}
	\author{Risheng Xu}
		\affiliation{%
			\institution{Kiel University}
			\institution{Mercedes-Benz AG}
			\city{Kiel \& Sindelfingen}
			\country{Germany}
		}
		\email{risheng.xu@mercedes-benz.com}
		\email{rxu@informatik.uni-kiel.de}
	
	\author{Philipp Sieweck}
		\affiliation{%
			\institution{Kiel University}
			\city{Kiel}
			\country{Germany}
		}
		\email{psi@informatik.uni-kiel.de}
	
	\author{Hermann von Hasseln}
		\affiliation{%
			\institution{Mercedes-Benz AG}
			\city{Sindelfingen}
			\country{Germany}
		}
		\email{hermann.v.hasseln@mercedes-benz.com}
	
	\author{Dirk Nowotka}
		\affiliation{%
			\institution{Kiel University}
			\city{Kiel}
			\country{Germany}
		}
		\email{dn@informatik.uni-kiel.de}
	%
	
	%
	%
	%


\begin{abstract}
We introduce \textbf{\texttt{preti}}, a novel framework for predicting software execution time during the early stages of development. \textbf{\texttt{preti}} leverages an LLVM-based simulation environment to extract timing-related runtime information, such as the count of executed LLVM IR instructions. This information, combined with historical execution time data, is utilized to train machine learning models for accurate time prediction. To further enhance prediction accuracy, our approach incorporates simulations of cache accesses and branch prediction. The evaluations on public benchmarks demonstrate that \textbf{\texttt{preti}} achieves an average Absolute Percentage Error (APE) of 11.98\%, surpassing state-of-the-art methods. These results underscore the effectiveness and efficiency of \textbf{\texttt{preti}} as a robust solution for early-stage timing analysis.
\end{abstract}


\maketitle
\section{Introduction}\label{sec:introduction}
In safety-critical real-time systems, timing analysis is essential to ensure that all tasks meet their deadlines. A widely adopted industry approach is measurement-based analysis, which helps mitigate the overly pessimistic estimates often associated with multi-core CPUs due to factors such as race conditions and cache coherency. Typically, timing measurements are conducted during the late stages of development, once all functional modules are integrated into a complete system. While this approach ensures that the system is evaluated under realistic conditions, it introduces two major challenges:

\begin{enumerate}
    \item \textbf{Failure Risks:} The measurement functionality itself, being part of the real-time system, is susceptible to failure if timing violations occur. For instance, the transmission of measurement results can be interrupted by an overrun task, leading to incomplete or inaccurate data. This not only compromises the reliability of the timing analysis but also increases the risk of undetected timing violations in the final system.
    
    \item \textbf{Limited Insight:} Unlike functional testing, which includes both system-level and unit-level tests, timing measurements are predominantly conducted at the system level. The absence of unit-level timing measurements restricts the ability to perform fine-grained debugging, making it difficult to identify and address performance bottlenecks early. As a result, timing issues often go undetected until system integration, where they escalate into more complex and costly problems.
\end{enumerate}

Despite the clear requirement of unit-level timing measurement before system integration, its practical implementation is hindered by challenge:

Unlike functional tests, which can be executed in simulators such as Software-in-the-Loop (SiL) \cite{demers2007generic}, timing measurements are typically performed on real embedded hardware. This is because timing behavior is highly dependent on hardware-specific factors such as CPU micro-architecture, memory hierarchy, and bus contention, which are difficult to accurately replicate in simulation environments. While simulators are effective for functional validation, they often fail to provide the level of accuracy and efficiency needed for timing analysis. As a result, real hardware measurements remain the preferred approach, despite their higher cost. The challenge are further exacerbated in large-scale software projects, where hundreds of engineers work in parallel. In such environments, the demand for frequent timing measurements—such as on every code commit—incurs substantial costs in terms of both hardware resources and engineering effort.

Recent advancements have explored the use of machine learning-based prediction models to reduce the reliance on frequent timing measurements. These models utilize features extracted from various representations of software, such as source code \cite{bonenfant2017early, huybrechts2018new, amalou2021we}, binary files \cite{amalou2024fast}, or assembly sequences \cite{amalou2022catreen, amalou2023cawet}, with time measurements collected from dedicated hardware serving as labels. While these approaches show promise, each feature extraction method has its limitations:

\begin{itemize}
    \item \textbf{Source Code:} Features derived from source code often fail to accurately reflect the actual execution behavior due to compiler optimizations. For example, dead code elimination, loop unrolling, and inlining can significantly alter the executed instructions compared to the original source code. As a result, features extracted from source code may overcomplicate the logic and yield overly pessimistic predictions.

    \item \textbf{Binary File:} Analyzing executed assembly sequences from binary files using static code analysis is challenging. While identifying instructions within a basic block is straightforward, linking these blocks to form the complete execution path triggered by specific input parameters is non-trivial. Traditional tools, such as Implicit Path Enumeration Technology (IPET) \cite{li1995performance}, enumerate possible paths without considering input-dependent behavior, which can lead to overestimations of path length. Consequently, binary-based analysis remains an incomplete solution for predicting execution time accurately.
    
    \item \textbf{Assembly Sequence:} Although executed assembly sequence contain precise execution information, capturing such sequences via hardware debugger (e.g. Lauterbach \cite{lauterbach_debugger}) is as costly as performing time measurements on most platforms. Additionally, parsing assembly sequence requires tools tailored to specific instruction sets, adding further complexity. As a result, assembly-based analysis is often impractical for early-stage timing prediction.
\end{itemize}

Given these limitations, there is a clear need for a hardware-independent yet representative data format to describe time-related features of function modules, making it applicable for early-stage prediction models. We propose that the LLVM Intermediate Representation (IR) \cite{lattner2004llvm} is a promising format for this purpose. LLVM IR provides a high-level, platform-agnostic representation of code that captures both the structure and semantics of the program while remaining independent of specific hardware architectures. This makes it an ideal candidate for extracting features that are both representative of execution behavior and suitable for machine learning-based timing prediction.

\section{Contributions}
In this work, we present \textbf{\texttt{preti}}, an end-to-end framework for predicting the software execution time under specific input parameters. Our contributions are highlighted as follows:

\begin{enumerate}
     \item We develop multiple LLVM passes to instrument the LLVM IR code, enabling the modeling of dynamic execution behavior. The instrumentation generates detailed execution traces, including basic block execution counts, data cache access hits and misses, and branch decisions. These traces capture the runtime behavior of the program, providing a robust foundation for accurate execution time prediction. 

    \item We process the trace data to extract meaningful feature vectors, which are used to train machine learning models. By leveraging existing timing measurements as labels, our models are trained on a dataset of \textbf{1348} code samples, all compiled with the \texttt{-O2} optimization level.    

    \item Our models achieve an Absolute Percentage Error (APE) of 11.98\% on the public benchmark, surpassing state-of-the-art approach. The results highlight \textbf{\texttt{preti}} as a practical and efficient solution for early-stage timing analysis, reducing the demand for time-consuming measurements.
\end{enumerate}

\section{Background}
\subsection{Use Case}
In industry, the development of real-time systems typically follows a structured process consisting of several stages: requirements analysis, module implementation, unit test with early-stage timing analysis (ESTA), scheduling and integration (S\&I), and system test with late-stage timing analysis (LSTA). Scheduling plays a critical role in this process, as it determines the allocation of tasks to CPUs and data to memory. We define the early stage as the phase before S\&I, during which software modules are developed and tested independently. The late stage begins after S\&I, when the entire system is integrated and available for testing.

The primary goal of LSTA is to ensure that the complete real-time system is safe and free from timing violations. In contrast, ESTA provides initial timing estimates for individual software modules during the early stage. These estimates serve two key purposes:
\begin{itemize}
    \item \textbf{Scheduling Support :} Scheduling and timing are interdependent. While scheduling decisions can alter timing behavior, they require initial timing data to begin. ESTA provides this critical input, enabling schedulers to make informed decisions about task allocation and resource management.

    \item \textbf{Debugging Aid :} If timing violations occur during LSTA, ESTA estimates serve as additional debugging information. This is particularly important when timing violations disrupt the LSTA module itself, rendering it non-debuggable.
\end{itemize}

Currently, ESTA relies on measurement-based approaches. However, the input values used for these measurements, originally designed for functional unit tests, may not cover the longest execution paths, such as reaching the maximum loop iteration. Therefore, additional effort is required to find input values that lead to the longest execution paths.

Heuristic optimization algorithms, like evolution control strategy \cite{mariani2013design} \cite{schwarzer2019compilation}, can solve this issue by exploring multiple solutions to find the optimal one. However, they require evaluations at each step, which increases the workload on the measurement system significantly. 

For example, if a project contains 1,000 function modules, each with one code commit per day and 1000 test cases are executed for heuristic algorithms to find the longest path, the required measurements can exceed one million per day. If we further consider the different CPU-data memory-program memory combinations (e.g., performance/efficient CPU, scratchpad/RAM), the measurement can be easily increased 10 times, i.e. ten million per day. The sheer volume of measurements far exceeds the capacity of most teams, highlighting the need for a more scalable and efficient approach.

With \textbf{\texttt{preti}}, we can quickly evaluate each step without requiring measurements, significantly reducing the measurement workload. The impact of prediction errors is mitigated by considering multiple candidates for the longest paths, with their actual lengths determined through targeted time measurements.

\subsection{aCET vs. wCET}
It is important to note that the predicted execution time in this work refers to the average core execution time (aCET) of a path from a software module running on a single-core CPU with a cold cache. This is distinct from the global worst-case core execution time (wCET), which involves complex multi-core interactions and race conditions.

The execution time of a software module is primarily determined by two factors: the triggered path length, and the resource scenario (e.g. CPU contention, memory allocation). Since the resource scenario is not yet determined in ESTA, our focus is on evaluating the path length to identify the longest path. In this context, aCET serves as a more suitable performance indicator than wCET. Without detailed resource information, wCET tends to be overly pessimistic, making it difficult to fairly compare execution paths and determine which one is longer.




\section{Related Work}\label{sec:rel_work} 
The existing literature relevant to this work can be primarily categorized into three domains: early-stage timing analysis, LLVM-based timing tools, and machine learning models for timing prediction.

\subsection{Early-Stage Timing Analysis} 
The concept of early-stage timing analysis was initially introduced by Altenbernd et al. in \cite{altenbernd2016early}. Our work builds upon this concept by translating source code into an intermediate representation and performing simulations to extract desired information. However, there are two key distinctions between our approach and previous work. Firstly, we utilize the LLVM IR instead of a virtual instruction set, which provides a more representative format of the source code. Secondly, we explicitly instrument the data cache accesses and branch choices during simulation, which are critical factors affecting timing performance.

Bate et al. \cite{bate2020establishing} introduced a series of fitness functions to evaluate the quality of time measurements. These fitness functions can also be applied in the early stage to select the potential longest execution path. Compared to this work, our framework offers a more straightforward approach by directly predicting the execution time using machine learning models.

\subsection{LLVM-based Timing Tools} 
Several tools have been developed using LLVM for timing analysis. Hahn et al. \cite{hahn2022llvmta} introduce LLVMTA, a comprehensive static wCET analysis tool that performs path, pipeline, and cache analysis on given C code. Touzeau et al. \cite{touzeau2023scalar} apply the LLVM ScalarEvolution pass to conduct symbolic analysis of the data cache. Both of these works focus on static analysis and aim to ensure that the results do not underestimate real-world scenarios. However, in the early stages of development, the lack of resource scenario often leads to over-pessimistic estimates. Therefore, we adopt an instrumentation-simulation approach instead of relying solely on static analysis.


\subsection{Machine Learning for Timing} 

The use of machine learning models for timing prediction was first proposed by Bonenfant et al. \cite{bonenfant2017early} and later implemented by Huybrechts et al. \cite{huybrechts2018new}, with further improvements by Amalou et al. in the WE-HML framework \cite{amalou2021we}. These studies derive feature vectors directly from the source code, excluding potential compiler optimizations. However, source code often includes dead code, constant expressions, and invariant code within loops that may be optimized by the compiler even at low optimization level (e.g. \texttt{-O0}), leading to unsatisfactory results.

The ITHEMAL framework developed by Mendis et al. \cite{mendis2019ithemal} treats the assembly sequence as a language sentence and employs the Long Short-Term Memory (LSTM) model for timing prediction. Amalou et al. further refined this model by incorporating execution context in CATREEN \cite{amalou2022catreen} and subsequently introduced the Transformer architecture in CAWET \cite{amalou2023cawet}. All three frameworks require assembly sequences as input to the machine learning model. However, in our use case, acquiring such sequences is as expensive as performing time measurements, making this approach less practical at the early stage.

The ORXESTRA \cite{amalou2024fast} by Amalou et al. and WORTEX \cite{reymond2024wortex} by Reymond et al. focus on predicting execution time at the basic-block level. In contrast, our approach focuses on extracting path-related information through simulation. These are two complementary approaches. Given sufficient training data, our approach could be combined with these machine learning architectures to further enhance prediction accuracy.

\section{Tool Architecture}\label{}
\subsection{Overview}

The \textbf{\texttt{preti}} framework comprises two main submodules: the \textbf{\texttt{preti\_llvm}} LLVM-based simulation and analysis framework, and the \textbf{\texttt{preti\_ml}} machine learning framework. Together, these components form an end-to-end solution for predicting the execution time of a given function written in C.

Given a function, we consider the execution time of this function is primarily determined by following factors:

\begin{enumerate}
    \item \textbf{Instruction Count:} The number of executed IR instructions, such as mathematical operations, variable initialization, and control flow instructions, directly impacts execution time. 
    
    \item \textbf{Data Access: } The memory hierarchy of CPUs introduces variability in data access times. Whether the accessed data resides in fast memory (e.g. cache or scratchpad) or slower memory (e.g. RAM) significantly affects execution time.

    \item \textbf{Instruction Fetch: } Similar to data access, the efficiency of fetching instructions depends on their allocation in memory. Instructions stored in fast memory result in faster execution, while those fetched from slower memory incur additional cycles.
    
    \item \textbf{Branch Prediction: }  Modern CPUs employ branch prediction to speculatively execute instructions ahead of time. If the prediction is correct (branch hit), the pipeline continues smoothly without stalls. However, if the prediction is incorrect (branch miss), the CPU must flush the pipeline and refetch the correct instructions, resulting in performance penalties.
\end{enumerate}

There are potentially other factors that may impact execution time as well, such as register spills. However, we do not observe such behavior on our experiment, so they are excluded from our model.

\begin{figure}[h]
	\centering
	\includegraphics[width=0.4\textwidth]{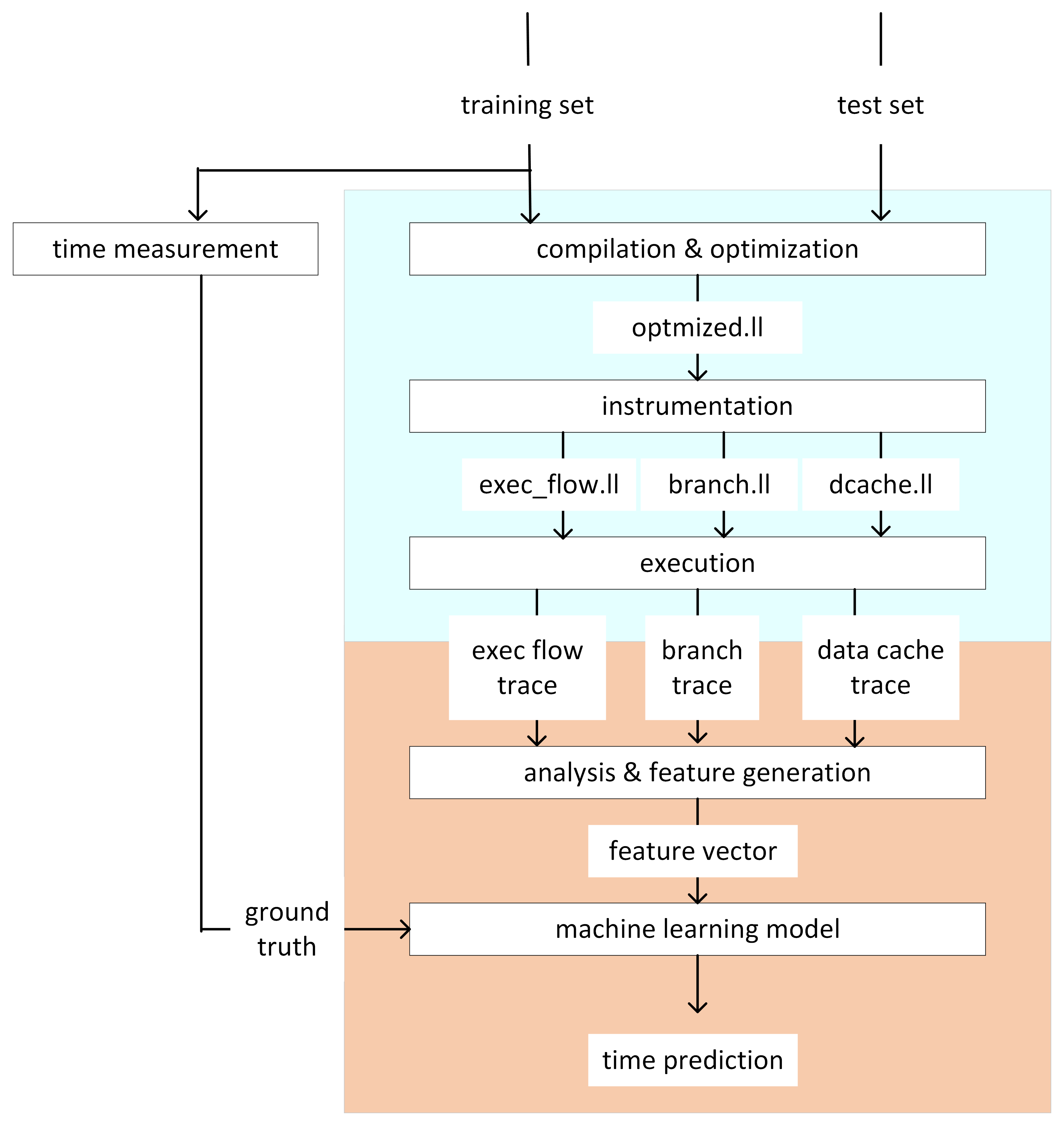}
	\caption{The Overview of \textbf{\texttt{preti}} framework}
	\label{fig:preti_overview}

\end{figure}

The workflow of \textbf{\texttt{preti}} is as shown in Figure \ref{fig:preti_overview}:

\begin{enumerate}
    \item \textbf{Code Compilation:} The C program is initially compiled into LLVM IR file \texttt{optimized.ll} using the Clang compiler. Various compiler optimizations are applied in this phase to ensure the IR accurately reflects real-world scenarios.
    
    
    \item \textbf{Instrumentation:} After compiling, the \texttt{optimized.ll} is instrumented to produce three distinct executable IR files: \newline \texttt{exec\_flow.ll}, \texttt{branch.ll}, and \texttt{dcache.ll}. Instrumentation involves inserting different types of counters and probes into the original functions to monitor various aspects of execution:
        \begin{itemize}
            \item \texttt{exec\_flow.ll}: Tracks the counts of each basic block is executed, as well as the data volume of initialized variables.
            \item \texttt{branch.ll}: Monitors branch predictions to capture information about instruction fetch behavior.
            \item \texttt{dcache.ll}: Simulates data cache accesses to track cache hits and misses.
        \end{itemize}
    
    \item \textbf{Simulation:} These instrumented IR files are then executed using the LLVM Interpreter \texttt{lli}. The execution generates traces that record the values of the inserted counters. These traces include detailed information about the execution flow, branch decision, and data cache accesses.
    
    \item \textbf{Feature Extraction:} The generated trace files are processed by the \texttt{preti\_ml} framework to extract feature vectors. These vectors represent time-related characteristics of the software.
    
    \item \textbf{Training:} Using the extracted features and historical time measurements as labels, the \texttt{preti\_ml} framework trains machine learning models to predict execution times.
    
    \item \textbf{Prediction:} Once trained, the machine learning models can predict the execution time of new C programs based on their LLVM IR representations and the features extracted during simulation.
\end{enumerate}

\subsection{Execution Flow Simulation}
The execution flow simulation is a critical component of \textbf{\texttt{preti\_llvm}}, designed to accurately model the execution behavior of a given software module with inputs. This simulation involves tracking the execution of basic blocks—straight-line sequences of instructions with no branches except at the entry and exit—and quantifying the number of instructions executed.

Each basic block is instrumented with three types of counters:

\textbf{Instruction Counter}: 
A counter is inserted at the entry point of every basic block. This counter is incremented each time the basic block is executed. By analyzing the number of executed basic blocks, we derive the number of different instructions executed during the program's run.

After the simulation, a simple instruction cache analysis is performed. Given that the size of our modules is typically much smaller than the instruction cache, we assume that all instructions start in a cold state (i.e., not cached) but remain always cached throughout the entire execution. The total number of cold instructions—those initially fetched from memory—is then used as a feature for the machine learning model.

\textbf{Call Counter}:
A counter is inserted before every call instruction to count the call to sub function, the instructions executed in sub function are then recursively added to the caller. 

\textbf{Memory Counter}: 
A counter is inserted before memory instructions, which are \texttt{memcpy}, \texttt{memset}, and \texttt{calloc}. This counter records the total data volume operated on by these memory instructions during execution. For example, if a \texttt{memcpy} instruction copies 100 bytes of data, the memory counter for \texttt{memcpy} increases by 100. 

We observe that the execution time of these IR instructions is linearly related to the data volume they handle. This is because, at the assembly level, these high-level IR operations are translated into multiple instructions, each of which processes a fixed amount of data.
 





\subsection{Data Cache Simulation}
Another critical component of \textbf{\texttt{preti\_llvm}} is the data cache simulation, aimed at accurately modeling the impact of data cache behavior on execution time. Modern CPUs typically employ a multi-level memory hierarchy, where the execution time of \texttt{load} and \texttt{store} instructions depends on whether the required data is already in the cache (cache hit) or not (cache miss). Accurately modeling this behavior is essential for realistic performance analysis.

The data cache exhibits both spatial locality (caching neighboring data) and temporal locality (caching recently accessed data). These behaviors are described using five key properties::
\begin{enumerate}
\item \textbf{Cache Size}: The total size of the data cache.
\item \textbf{Line Size}: The amount of data that is transferred to the cache in a single \texttt{load} or \texttt{store} instruction.
\item \textbf{Associativity}: The number of potential cache lines an incoming block of data can occupy.
\item \textbf{Replace Policy}: The policy used to replace existing data with new data.
\item \textbf{Write Policy}: The timing of when to write data in the cache back to main memory.
\end{enumerate}

According to our CPU, the Infineon Aurix TC387, we have built a data cache model with the following specifications: 16 kB cache size, 256 bits line size, 2-way set associative, write-back strategy, and least-recent-used (LRU) replacement. That is, when the CPU requires data, it transfers the data, along with its neighboring data, with a total size of 256 bits, into the cache. This data set may be placed in two cache lines, and the least recently used data is replaced.

When the CPU writes data to the cache, it only updates the cache copy. The modified data is not immediately written back to main memory. Instead, the cache controller stores the modified data in the cache and marks it as "dirty." When the cache line is eventually replaced, the dirty data is written back to main memory.

We insert a initialization function of cache model at the beginning of \texttt{main} function, and update the cache status after every \texttt{load} and \texttt{store} instruction.



Once the \texttt{main} function returns, the cache model is reset to initial status, meanwhile a C function is called to flush the data cache of the CPU and guarantee the cache is cold for next execution.

The memory allocation of data in an integrated system may differ from that in an LLVM-based simulation. This is because the final memory layout is determined after the execution of \textbf{\texttt{preti}}, creating a cyclic dependency on \textbf{\texttt{preti}}’s output. Despite these differences in memory allocation, the following cache behaviors remain consistent between simulation and real hardware:

\begin{enumerate}
    \item \textbf{Temporal Locality:} Both single variables and array elements remain cached if they have been accessed recently. This behavior is preserved regardless of the memory layout, as temporal locality depends on access frequency rather than memory arrangement.
    
    \item \textbf{Spatial Locality for Arrays:} When an element from an array is accessed, its neighboring elements are also cached, as arrays in C are allocated in contiguous memory blocks. This ensures that spatial locality for arrays is maintained in both simulation and real hardware.
\end{enumerate}

 A potential discrepancy arises when considering the spatial locality of single variables (e.g., integers). In real hardware, a single variable’s neighbors in memory may differ from those in the simulation due to the differing memory layouts. This could lead to errors when predicting whether a single variable remains cached based on the access of adjacent variables. However, in C, caching for single variables is primarily governed by temporal locality rather than spatial locality, so the impact of this error is minimal.


    



\subsection{Branch Simulation}
Lastly, branch prediction is a critical factor in simulation because modern CPUs rely on multi-stage pipelines to prefetch instructions. When a conditional branch (e.g., a \texttt{br} instruction) is encountered, the CPU predicts the likely path of execution and prefetches the corresponding instructions. If the prediction is incorrect, the pipeline must discard the prefetched instructions and refetch the correct ones, resulting in stall cycles.

To evaluate branch prediction difficulty, we implemented a 2-bit branch predictor to predict the outcomes of all branch instructions. The prediction is then compared with the value of decision variable and the \texttt{br\_hit} and \texttt{br\_miss} are then counted. This is done by inserting a function call before each conditional branch instruction. The function both predicts the branch decision and updates the internal predictor state.

To evaluate the difficulty of branch prediction, we implemented a 2-bit branch predictor to predict the outcomes of all branch instructions. This predictor works by comparing its predictions with the actual value of the decision variable, and then counting the number of \texttt{br\_hit} (correct predictions) and \texttt{br\_miss} (incorrect predictions). To achieve this, we inserted a function call before each conditional branch instruction. This function predicts the branch decision and updates the internal state of the predictor based on the actual outcome.

The 2-bit branch predictor maintains four states and updates them based on whether the prediction matches the actual outcome:

\begin{enumerate}
    \item \textbf{Strongly Taken (ST):} Predicts "taken."
    \begin{itemize}
        \item If the branch is taken, remain in ST.
        \item If the branch is not taken, transition to WT.
    \end{itemize}

    \item \textbf{Weakly Taken (WT):} Predicts "taken," but close to switching.
    \begin{itemize}
        \item If the branch is taken, transition to ST.
        \item If the branch is not taken, transition to WNT.
    \end{itemize}

    \item \textbf{Weakly Not Taken (WNT):} Predicts "not taken," but close to switching.
    \begin{itemize}
        \item If the branch is not taken, transition to SNT.
        \item If the branch is taken, transition to WT.
    \end{itemize}

    \item \textbf{Strongly Not Taken (SNT):} Predicts "not taken."
    \begin{itemize}
        \item If the branch is not taken, remain in SNT.
        \item If the branch is taken, transition to WNT.
    \end{itemize}
\end{enumerate}

This branch prediction mechanism helps estimate the impact of mispredictions on execution time and improves the accuracy of our simulation. Upon returning of main function, the function resets the branch model and prints the counts of predictions (hit, miss) to the branch trace file for further analysis. 

Additionally, we observe that a cyclic execution of a long sequence of basic blocks, such as \( A \rightarrow B \rightarrow C \rightarrow A \rightarrow B \rightarrow C \), is significantly slower than a simple loop over a single basic block, such as \( A \rightarrow A \rightarrow A \rightarrow A \rightarrow A \rightarrow A \), even when the total number of executed instructions are close and branch prediction is always correct. We hypothesize that this performance discrepancy arises from micro-architectural inefficiencies inherent in transitioning between basic blocks. Specifically, the hardware incurs stall cycles whenever execution jumps to a different basic block, even if the branch itself is perfectly predicted.

To account for this behavior, we incorporate the count of basic block jumps \texttt{bb\_jump} as a feature in our machine learning model. This feature captures the frequency of transitions between basic blocks, providing a proxy for the associated micro-architectural overhead.

\subsection{Machine Learning Model}
The collected time measurement data is valid only for a specific hardware with particular compiler configuration, meaning there is no publicly available large dataset. As a result, training data is typically gathered shortly before the training phase in most use cases. Due to measurement limitations, we often have a limited amount of training data. Therefore, we primarily focus on using simple yet robust machine learning models to minimize the risk of overfitting.

After a preliminary investigation, we identified the following models as robust against small datasets: linear regression (LR),  Huber regression (HR), random forest (RF), and multi-layer perception (MLP).

We evaluate these models using two metrics to gain a comprehensive understanding of their performance: the average percentage error (APE) and the symmetric average percentage error (sAPE).


For a given sample $n$ from a data set $N$, these two metrics are defined as follows:
\begin{equation}
\text{sAPE}(n) = \left| \frac{A(n) -P(n)}{A(n) + P(n)} \right| \times 200\%
\end{equation}

\begin{equation}
\text{APE}(n) = \left| \frac{A(n) - P(n)}{A(n)} \right| \times 100\%
\end{equation}

where:
\begin{itemize}
    \item \( P(n) \) is the predicted value,
    \item \( A(n) \) is the actual value.
\end{itemize}

While random forest does not have an explicit loss function, the other models (LR, HR, and MLP) are trained using loss functions based on mean squared error (MSE) or its variants. Although the evaluation metrics are based on APE or sAPE, MSE-based loss functions are used during training for the following reasons:

\begin{enumerate}
    \item \textbf{Sensitivity to Small Values}: Both sAPE and APE involve division by the true value \( A(n) \). If \( A(n) \) is very small or close to zero, these metrics can explode to very large values or become undefined (division by zero). This is particularly problematic for short execution times, where some background measurement error (e.g., due to OS activity) can dominate the true execution time. As a result, short-execution-time measurements, which are less reliable, can disproportionately influence the training when using APE or sAPE as loss functions.
    
    \item \textbf{Non-Smooth Gradients}: the division, together with absolute operation, also makes their gradients non-smooth near zero. This can lead to unstable training dynamics, especially for gradient-based optimization methods.
\end{enumerate}

Let $d(n) = (A(n) - P(n))$, the MSE is used as the loss function of LR and MLP:
\begin{equation}
   L^{nn}(N) = L^{lr}(N) = \frac{1}{|N|} \sum_{n=1}^{n} d(n)^2
\end{equation}

For HR, the loss function is defined as:
\begin{equation}
L^{hr}(N,\delta) = 
\begin{cases} 
\frac{1}{|N|} \cdot \sum_{n \in N} \frac{1}{2} d(n) ^2 & \text{if} |d(n)| \leq \delta, \\

\frac{1}{|N|} \cdot \sum_{n \in N} (\delta \cdot |d(n)| - \frac{1}{2} \delta^2)  & \text{otherwise}.

\end{cases}
\end{equation}

where 
$\delta$ is the threshold coefficient.

The Huber loss function combines two loss functions: it behaves as a linear loss when the absolute difference between the true and predicted values exceeds a threshold $\delta$, and as a squared loss otherwise. This design makes the Huber loss more robust to outliers in the dataset, as it reduces the impact of extreme values on the overall loss.
\section{Experiment}\label{sec:experiment}
In this section, we describe the experimental setup and methodology used to train and evaluate our prediction model. Specifically, we detail the dataset used for training and testing in Sections \ref{sec:dataset}, the measurement setup in \ref{sec:meas}, tool setup in Section \ref{sec:tool}, the compiler configurations in Section \ref{sec:compiler}, the hyperparameters of the model in Section \ref{sec:hyperparam}, and the feature engineering process in Section \ref{sec:feature}. Finally, we present the evaluation of our model in Section \ref{sec:result} and discuss the error in Section \ref{sec:error}, the overhead in Section \ref{sec:overhead}.


\subsection{Dataset}\label{sec:dataset}
The \textbf{\texttt{preti\_ml}} framework employs a supervised machine learning approach, utilizing training code and its corresponding execution time measurements to train our model.

When constructing the training set, we focus on two key properties:
\begin{enumerate}
    \item \textbf{Representativeness}: 
    The training data must accurately reflect the characteristics of real-world code. To achieve this, we collect 60 popular algorithms and implement 766 test cases, forming the \texttt{real\_world} training set. This ensures that the model is exposed to a diverse range of code patterns and behaviors commonly encountered in practice.

    \item \textbf{Diversity}: 
    Real-world code often exhibits an imbalanced usage of different C operators. For instance, operators like addition and assignment are frequently used, while division and multiplication are relatively rare. Training the model solely on real-world code would require an impractically large number of samples to capture the impact of these less frequent operators. To address this, we create a \texttt{simple\_loop} training set, where specific operators (e.g., division) are iteratively executed within loop structures. This increases the frequency of these operators in the training data, enabling the model to learn their effects more efficiently. Additionally, we include sanity checks to exclude illegal operations, such as division by zero, ensuring the validity of the training data. The \texttt{simple\_loop} set contains 568 test cases.    
\end{enumerate}

For the test set, we use the same test cases as CATREEN \cite{amalou2022catreen} to ensure a fair comparison. CATREEN is one of the latest works focusing on predicting the execution time of triggered paths within software functions. 19 algorithms from CATREEN are evaluated in the comparison. There are also 5 algorithms excluded due to two primary reasons. First, the memory requirements for their execution exceed the capacity of our embedded hardware. Second, these algorithms rely on functions from C standard library, such as \texttt{fopen}, that are not currently modeled in \texttt{\textbf{preti}}. The integration of such functions remains an area for future work.

\begin{figure}[h]
    \centering
    \includegraphics[width=0.55\textwidth]{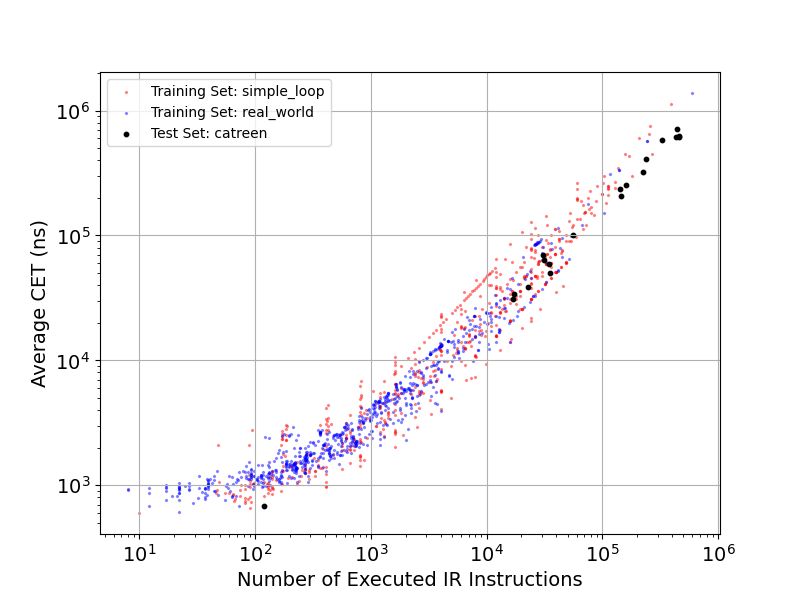}
    \caption{Execution Time vs. Number of Executed IR Instructions}
    \label{fig:cet_vs_llvm}
\end{figure}

As illustrated in Figure \ref{fig:cet_vs_llvm}, our dataset reveals a strong positive correlation between measured execution time and the number of executed LLVM instructions, demonstrating the effectiveness of our simulation. The \texttt{simple\_loop} set exhibits a broader distribution compared to the \texttt{real\_world} set, reflecting its greater diversity as discussed above.





\subsection{Measurement Configuration}\label{sec:meas}
We use the Gliwa \cite{gliwa} Processor-in-the-Loop (PiL) system to measure the execution time of the training and test sets. Initially, the code is wrapped by the Gliwa OS and flashed onto an Infineon evaluation board equipped with an Aurix TC387 CPU. This CPU is a superscalar multi-core processor featuring three in-order execution pipelines and branch prediction. During the measurements, only CPU0 is utilized, with both the data and program caches enabled. The caches are reset after each test case to ensure consistent measurement conditions.

Regarding the memory configuration, the measured code is flashed to the program flash memory (PFLASH0). The stack is allocated on the data scratchpad (DSPR0), the heap is placed on the local memory unit (LMU0), and global variables are stored in the distributed local memory (dLMU0), which offers performance comparable to the data scratchpad.

\subsection{Simulation and Training}\label{sec:tool}
For simulation and machine learning training, we use a Docker container running on a ThinkPad laptop equipped with a 12th Gen Intel(R) Core(TM) i7-1280P processor clocked at 1.80 GHz. The container is allocated 20 CPU cores and 2 GB of RAM, with Ubuntu 20.04 as the operating system.

The simulation environment is built using LLVM and Clang 10 \cite{lattner2004llvm}. For the machine learning model, we employ the PyTorch 2.0.1 \cite{paszke2019pytorch} and Scikit-learn 1.4.2 \cite{pedregosa2011scikit} frameworks.

\subsection{Compiler Optimization}\label{sec:compiler}
To compile the C code for our measurements, we use the Tasking Compiler \cite{tasking} 6.2.2. We enable the default optimization level \textbf{\texttt{-O2}}, which corresponds to the following individual optimization flags:

\begin{itemize}
    \item \textbf{\texttt{-OacefgIklMnoprsUvwy}}
\end{itemize}

This configuration enables all common optimizations except those that may increase the size of the binary file, which aligns with our real-world use case constraints.

To ensure consistency between the Tasking compiler and the LLVM-based simulation, we apply similar optimizations in Clang. Specifically, we use the \textbf{\texttt{-O2}} optimization level in Clang, with the following additional options disabled:

\begin{itemize}
    \item \textbf{\texttt{-fno-unroll-loops}}: Disables loop unrolling
    \item \textbf{\texttt{-fno-fast-math}}: Disables aggressive floating-point optimizations
    \item \textbf{\texttt{-fno-inline}}: Disables function inlining
    \item \textbf{\texttt{-fno-vectorize}}: Disables automatic vectorization
\end{itemize}

\subsection{Hyperparameters}\label{sec:hyperparam}
To mitigate the overfitting problem inherent in small datasets, we employ multiple simple regression models, each providing a distinct perspective on data analysis. This approach enhances the robustness and generalizability of our findings. By comparing the performance of these models, we gain deeper insights into the underlying patterns within the dataset, thereby improving the reliability of our predictions.

We use the Scikit-learn framework to train the Linear Regression (LR), Huber Regression (HR), and Random Forest (RF) models. While LR does not require specific hyperparameters, HR and RF are trained using the default hyperparameters provided by Scikit-learn. The parameters are listed in Table \ref{tab:hyper}. For the Multi-Layer Perceptron (MLP) model, we use PyTorch and leverage the Optuna framework \cite{akiba2019optuna} to tune the hyperparameters for optimal performance.

\begin{table}[h]
\caption{Hyperparameters}
\begin{tabular}{|ll|ll|ll|}
\hline
\multicolumn{2}{|l|}{HR}                   & \multicolumn{2}{l|}{MLP}             & \multicolumn{2}{l|}{RF}                 \\ \hline
\multicolumn{1}{|l|}{$\epsilon$}             & 1.35   & \multicolumn{1}{l|}{$\alpha$}  & 2e-5   & \multicolumn{1}{l|}{n}           & 100  \\ \hline
\multicolumn{1}{|l|}{max iter.}             & 100      & \multicolumn{1}{l|}{batch}  & 4      & \multicolumn{1}{l|}{max depth}   & 64   \\ \hline
\multicolumn{1}{|l|}{L2 reg.}             & 0.0001     & \multicolumn{1}{l|}{epoch}  & 10     & \multicolumn{1}{l|}{min split}   & 2   \\ \hline
\multicolumn{1}{|l|}{}   &      & \multicolumn{1}{l|}{$\lambda$} & 0.0001 & \multicolumn{1}{l|}{min leaf}    & 1    \\ \hline
\multicolumn{1}{|l|}{} &  & \multicolumn{1}{l|}{}       &        & \multicolumn{1}{l|}{max feature} & 1 \\ \hline
\end{tabular}
\label{tab:hyper}
\end{table}

\subsection{Feature Generation}\label{sec:feature}
In previous work, two types of features are primarily used: non-sequential and sequential. Non-sequential models \cite{amalou2021we} \cite{huybrechts2018new} utilize the count of different instructions or operators as features. In contrast, sequential models \cite{amalou2022catreen} \cite{amalou2023cawet} \cite{amalou2024fast} treat the executed assembly sequence as sentences in a natural language, using them as feature vectors for the model. Theoretically, sequential models are more representative as they capture not only the number of executed instructions but also their execution order, which implicitly reflects the CPU pipeline state. However, sequential models, such as Recurrent Neural Networks (RNNs) or Transformers, require significantly more diverse data than non-sequential models. This poses a challenge in industrial use cases due to the high cost of time measurement.

In this work, we adopt the concept of non-sequential models and use the count of executed LLVM IR instructions as input to the model. Additionally, we refine the classification of certain instructions based on simulation results. Specifically, we distinguish between different types of \texttt{load} and \texttt{store} instructions by categorizing them as \texttt{load\_hit}, \texttt{load\_miss}, \texttt{store\_hit}, and \texttt{store\_miss}. Furthermore, we classify conditional branch instructions (\texttt{br}) into \texttt{br\_miss} and \texttt{br\_hit} based on branch simulation results. We also count basic block jumps \texttt{bb\_jump} as additional information, i.e., when the next executed basic block differs from the previous one. Lastly, we count the number of cold-started instructions as \texttt{inst\_miss} to indicate the performance of the instruction cache.

For memory-related functions such as \texttt{memset}, \texttt{memcpy}, and \texttt{calloc}, their values are determined by the data volume they operate on.

After excluding instructions that do not appear in our use case, we consider the following LLVM instructions, listed in Table \ref{tab:feature_inst}, as feature inputs to the machine learning model:

\begin{table}[h]
\centering
\caption{Feature Instructions}
\begin{tabular}{>{\ttfamily}l >{\ttfamily}l >{\ttfamily}l >{\ttfamily}l}
\toprule
add & fadd & sub & fsub \\
and & or & xor & shl \\
lshr & ashr & icmp & fcmp \\
zext & sext & fptosi & uitofp \\
sitofp & fneg & sdiv & fdiv \\
mul & udiv & urem & fmul \\
srem & br\_hit & br\_miss & br\_uncond \\
store\_miss & store\_hit & load\_miss & load\_hit \\
switch & getelementptr & phi & alloca \\
memset & memcpy & calloc & malloc \\
inst\_miss & bb\_jump &  & \\
\bottomrule
\end{tabular}
\label{tab:feature_inst}
\end{table}

This detailed classification enhances the granularity of the input features, leading to more accurate performance by better reflecting the runtime behavior.

\subsection{Performance Analysis}\label{sec:result}
This section evaluates four models: Linear Regression (LR), Huber Regression (HR), Multilayer Perceptron (MLP), and Random Forest (RF). Each model is evaluated with Symmetric Absolute Percentage Error (sAPE) and Absolute Percentage Error (APE). The average prediction errors are listed in Table \ref{tab:average_losses}. Detailed errors are provided in the Appendix due to page limitations.

According to the results in Table \ref{tab:average_losses}, the sAPE of all models ranges from 12.11\% to 21.59\%, and the APE ranges from 11.98\% to 18.63\%. RF achieves the best results. This can be attributed to its ability to strike a balance between flexibility and generalization. Compared to MLP, RF is more resistant to overfitting, particularly in small datasets. In contrast to LR and HR, RF is more flexible and capable of capturing nonlinear relationships in the data.

The relatively consistent performance of all models (with differences of less than 10\%) further highlights the effectiveness of our simulation, which generates high-quality feature vectors that are well-suited for all models. 

In Table \ref{tab:rf_catreen_itheml_comparison}, we compare our results with the performance of CATREEN and ITHEMAL from \cite{amalou2022catreen} in 19 algorithms. The RF model from \texttt{\textbf{preti\_ml}} achieves the best performance in 13 algorithms, with a 5.43\% lower average APE. Compared to CATREEN and ITHEMAL, which use traces captured by a hardware debugger, our approach uses LLVM traces from simulation, which are much easier to acquire, yet still achieves better performance.


\begin{table}[h!]
\centering
\begin{tabular}{lcccc}
\hline
\textbf{Metric} & \textbf{HR} & \textbf{LR} & \textbf{RF} & \textbf{NN} \\
\hline
sAPE & 13.26\% & 21.59\% & 12.11\% & 16.94\% \\
APE  & 13.09\% & 18.63\% & 11.98\% & 13.19\% \\
\hline
\end{tabular}
\caption{Average sAPE and APE for HR, LR, RF, and MLP on CATREEN Test Set.}
\label{tab:average_losses}
\end{table}

\begin{table}[h!]
\centering
\begin{tabular}{lccc}
\hline
\textbf{Algorithm} & \textbf{RF} & \textbf{CATREEN} & \textbf{ITHEMAL} \\
\hline
2mm             & 26.61\% & \textbf{17.2\%}  & 21.3\%  \\
3mm             & \textbf{1.49\%}  & 17.7\%  & 21.3\%  \\
atax            & \textbf{3.70\%}  & 23.2\%  & 24.8\%  \\
bicg            & 28.21\% & \textbf{18.7\%}  & 20.2\%  \\
bitcount        & 39.66\% & \textbf{9.3\%}   & 29.3\%  \\
covariance      & \textbf{7.61\%}  & 15.8\%  & 18.4\%  \\
dotigen         & 28.93\% & \textbf{21.5\%}  & 25.5\%  \\
floyd\_warshall & 16.95\% & \textbf{11.5\%}  & 21.6\%  \\
gemm            & \textbf{4.99\%}  & 19.8\%  & 21.1\%  \\
gemver          & \textbf{11.51\%}  & 20.9\%  & 22.5\%  \\
gesummv         & 12.74\% & \textbf{15.0\%}  & 15.3\%  \\
gramschmidt     & \textbf{3.79\%}  & 16.9\%  & 20.1\%  \\
lu              & \textbf{3.75\%}  & 18.7\%  & 22.1\%  \\
ludcmp          & \textbf{1.68\%}  & 11.0\%  & 16.9\%  \\
mvt             & \textbf{4.96\%}  & 25.7\%  & 25.3\%  \\
symm            & \textbf{0.89\%}  & 17.8\%  & 18.4\%  \\
syrk            & \textbf{4.72\%}  & 18.3\%  & 18.6\%  \\
trisolv         & \textbf{3.55\%}  & 9.1\%   & 11.9\%  \\
trmm            & \textbf{21.84\%} & 22.6\%  & 24.8\%  \\
\hline
\textbf{Average} & \textbf{11.98\%} & 17.41\% & 21.02\% \\
\hline
\end{tabular}
\caption{Comparison of RF, CATREEN, and ITHEMAL based on APE.}
\label{tab:rf_catreen_itheml_comparison}
\end{table}

\subsection{Error Analysis}\label{sec:error}
The prediction errors can be primarily attributed to the following three factors:

\textbf{Data Bias:} While we collect two datasets to ensure representativeness and diversity, the limited quantity of data remains a challenge for building a fully generalizable model. However, in real-world iterative software development, this limitation is less problematic. By training the model on historical code from previous or dependent projects and applying it to predict the future updates, the impact of data bias can be significantly mitigated. This approach ensures that the training and prediction code are closely aligned, reducing the risk of generalization errors.

\textbf{Hardware Microstructure:} The micro-architectural behavior of hardware cannot be fully captured through LLVM simulation. For instance, stall cycles introduced in the pipeline or memory bus are abstracted away in our current modeling. A more detailed modeling of these structures, as suggested by \cite{bai2023computing}, could provide benefits in the future.

\textbf{Measurement Noise:} Our measurement module is wrapped by the operating system, which introduces uncertain time costs due to OS activities such as task dispatch and interrupt handling.

\subsection{Overhead}\label{sec:overhead}
For a given functional module, depending on its complexity, it typically requires 100 to 500 test cases. Our time measurement system takes approximately 2 to 5 minutes to complete the entire measurement, which includes 5 steps:
\begin{itemize}
    \item Wrapper code generation
    \item Compilation
    \item Flashing
    \item Measurement
    \item Report generation
\end{itemize}

In contrast, the \texttt{preti} framework completes the simulation and prediction process in just 5 seconds, making it approximately 50 times faster than our current time measurement system. By leveraging \texttt{preti}, the measurement demand can be drastically reduced—from millions to just thousands per day. This remarkable improvement in efficiency not only accelerates the development process but also enhances the reliability of real-time systems.

\subsection{Limitations and Future Work}
Currently, \textbf{\texttt{preti}} only considers common LLVM IR instructions as features and does not yet support many functions from C standard library. In the future, we plan to extend \textbf{\texttt{preti}} by incorporating common C standard library functions, thereby enhancing its applicability to a broader range of use cases.

\section{Conclusion}\label{sec:concusion}
We present \textbf{\texttt{preti}}, the LLVM-based simulation and prediction framework for early-stage timing analysis. \textbf{\texttt{preti}} instrument the LLVM IR code and perform simulation with \texttt{lli}, then utilize the runtime information from simulation to predict the execution time. We evaluate the \textbf{\texttt{preti}} on the public benchmark and achieve 11.98\% APE and 12.11\% APE, surpassing state-of-the-art methods.

\bibliographystyle{ACM-Reference-Format}
\bibliography{main}

\onecolumn
\section{Appendix}

\begin{table}[h!]
\centering
\begin{tabular}{lcccccccc}
\hline
\textbf{Algorithm} & \multicolumn{2}{c}{\textbf{HR}} & \multicolumn{2}{c}{\textbf{LR}} & \multicolumn{2}{c}{\textbf{RF}} & \multicolumn{2}{c}{\textbf{NN}} \\
 & sAPE & APE & sAPE & APE & sAPE & APE & sAPE & APE \\
\hline
2mm & 18.48\% & 16.92\% & 47.91\% & 38.65\% & 21.05\% & 26.61\% & 10.22\% & 4.80\% \\
3mm & 18.26\% & 16.73\% & 50.95\% & 40.61\% & 0.66\% & 1.49\% & 10.14\% & 5.05\% \\
atax & 9.10\% & 9.54\% & 6.66\% & 6.45\% & 7.73\% & 3.70\% & 18.02\% & 9.66\% \\
bicg & 1.86\% & 1.84\% & 34.30\% & 29.28\% & 26.83\% & 28.21\% & 16.16\% & 16.25\% \\
bitcount & 39.60\% & 49.38\% & 41.01\% & 34.03\% & 51.48\% & 39.66\% & 99.30\% & 67.37\% \\
covariance & 0.82\% & 0.82\% & 8.59\% & 8.24\% & 6.40\% & 7.61\% & 0.58\% & 6.76\% \\
dotigen & 17.96\% & 16.48\% & 21.26\% & 19.22\% & 28.68\% & 28.93\% & 3.29\% & 7.86\% \\
floyd\_warshall & 14.33\% & 13.37\% & 10.84\% & 10.29\% & 12.94\% & 16.95\% & 15.41\% & 15.38\% \\
gemm & 7.90\% & 7.60\% & 7.29\% & 7.03\% & 4.37\% & 4.99\% & 4.41\% & 3.89\% \\
gemver & 8.81\% & 9.22\% & 10.49\% & 9.97\% & 8.90\% & 11.51\% & 29.62\% & 19.46\% \\
gesummv & 18.70\% & 20.63\% & 2.37\% & 2.34\% & 17.83\% & 12.74\% & 37.26\% & 25.15\% \\
gramschmidt & 31.74\% & 27.39\% & 38.36\% & 32.18\% & 6.27\% & 3.79\% & 18.15\% & 14.08\% \\
lu & 26.36\% & 23.29\% & 31.78\% & 27.42\% & 4.08\% & 3.75\% & 0.77\% & 2.84\% \\
ludcmp & 20.57\% & 18.65\% & 32.12\% & 27.68\% & 4.89\% & 1.68\% & 11.42\% & 8.77\% \\
mvt & 0.25\% & 0.25\% & 21.35\% & 19.29\% & 1.82\% & 4.96\% & 1.22\% & 10.24\% \\
symm & 0.52\% & 0.52\% & 3.89\% & 3.96\% & 0.76\% & 0.89\% & 30.01\% & 18.40\% \\
syrk & 10.00\% & 9.52\% & 9.16\% & 8.76\% & 2.73\% & 4.72\% & 3.47\% & 4.79\% \\
trisolv & 1.95\% & 1.97\% & 27.65\% & 24.29\% & 6.58\% & 3.55\% & 4.10\% & 9.79\% \\
trmm & 4.74\% & 4.63\% & 4.33\% & 4.24\% & 16.19\% & 21.84\% & 8.23\% & 0.07\% \\
\hline
\textbf{Average} & 13.26\% & 13.09\% & 21.59\% & 18.63\% & 12.11\% & 11.98\% & 16.94\% & 13.19\% \\
\hline
\end{tabular}
\caption{sAPE and APE metrics for test cases from CATREEN.}
\label{tab:algorithm_metrics_with_avg}
\end{table}
\end{document}